\documentstyle[epsf,twoside,fleqn,espcrc2]{article}

\newcommand{\eq}{\begin{equation}}
\newcommand{\en}{\end{equation}}
\title{First-Order Signals in Compact QED with Monopole Suppressed 
        Boundaries\thanks{Work supported by the Deutsche 
        Forschungsgemeinschaft under grant No.~Schi 257/3-2, Schi 257/1-4
        and by EC contract  CHRX-CT92-0051.}}
\author{\underline{Th.~Lippert}$^{\mbox{a}}$,
        A.~Bode$^{\mbox{a}}$, 
        V.~Bornyakov$^{\mbox{b}}$, 
        and K.~Schilling$^{\mbox{a,c}}$\\[8pt]
       {$^{\mbox{a}}$Department of Physics, 
        University of Wuppertal, D-42097 Wuppertal, Germany}\\
       {$^{\mbox{b}}$Institute of High Energy Physics, 
        142284 Protvino, Russia}\\
       {$^{\mbox{c}}$HLRZ c/0 KFA, D-52425 J\"ulich, Germany}}%

\begin{document}
\begin{abstract}
Pure gauge compact QED on hypercubic lattices is considered with
periodically closed monopole currents suppressed.  We compute
observables on sublattices which are nested around the centre of the
lattice in order to locate regions where translation symmetry is
approximately recovered. Our Monte Carlo simulations on
$24^4$-lattices give indications for a first-order nature of the U(1)
phase transition.
\end{abstract}
\maketitle
\setcounter{footnote}{1}

\section{INTRODUCTION}

The presence of metastabilities at the phase transition of compact QED
became obvious in simulations on $6^4$-lattices with periodic boundary
conditions (p.b.c.)\cite{JERSAK}.  With increasing lattice size, there
was growing confidence that the U(1) phase transition is exhibiting a
weakly first-order nature, culminating in the observation of extremely
strong hysteresis effects appearing in Monte Carlo series on
$16^4$-lattices\cite{LIPPERT1}.

This widely accepted opinion has been questioned recently.  The
authors of Refs.\ \cite{LANG1,LANG2} have implemented compact U(1) on
a lattice with trivial homotopy group, {\em i.e.}\ $SH_4$, the surface
of $H_5$.  No indications for metastabilities have been found at
$\beta_c$ in simulations on lattices of different sizes---up to a
number of degrees of freedom equivalent to a $16^4$-lattice with
hyper-toroidal topology. Similarly, fixed boundary conditions (f.b.c.)
hint at a continuous transition\cite{BAIG,SCHILLING}.

The implementation of U(1) on $SH_4$ as well as with f.b.c.\
explicitly breaks the translation symmetry and produces inhomogenities
of the free energy.  One might argue that the absence of first-order
effects can be explained in these terms.  Alternatively it has been
suggested in Ref.~\cite{LANG2} that the existence of topologically
non-trivial monopole loops wrapping around the lattice due to p.b.c.\
might lead to a first-order signature at $\beta_c$.

We can check this assumption by studying the model with p.b.c.\ but
periodically closed monopole loops forbidden and looking for
first-order signals.  For this purpose, we suppress monopole currents
emerging in the Monte Carlo update that would point outside the
boundary shell and thus would be closed due to p.b.c.

As monitoring quantity we measure the average plaquette on the
hyper-surface of subcubes nested around the centre of the lattice and
we try to establish a region where translation symmetry is recovered
approximately.

\section{SUPPRESSION OF MONOPOLES ON BOUNDARY}

\begin{figure}
\begin{minipage}[t]{3.5cm}
\epsfxsize=3.4 cm
\centerline{\epsfbox{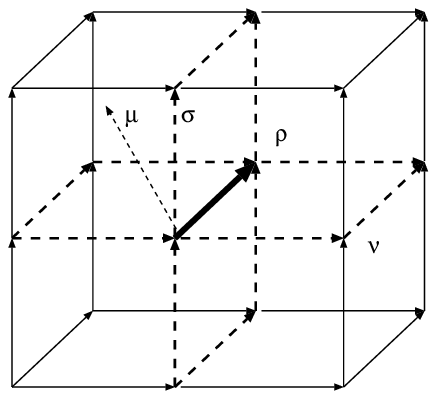}}
\vglue -12pt
\caption[a]{Plaquettes contributing to $m_{\mu}(x)$.\label{CUBE}}
\end{minipage}
\hfill
\begin{minipage}[t]{3.5cm}
\epsfxsize=3.1 cm
\centerline{\epsfbox{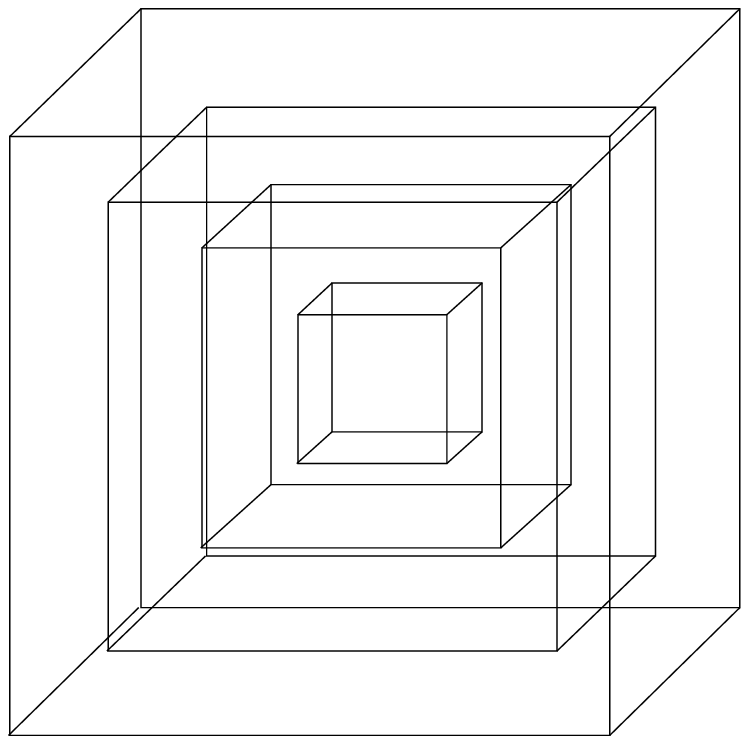}}
\vglue -12pt
\caption[a]{\sloppy Mea\-sure\-ments on nested shells.\label{NEST}}
\end{minipage}
\end{figure}

We suppress monopoles on the boundaries in such a way that spinwaves can
penetrate them.
The monopole current is defined by\cite{DEGRAND}
\begin{eqnarray}
m_{\mu}(x)&=& 
\frac{1}{2}\epsilon_{\mu\nu\rho\sigma}\Big[
n_{\rho\sigma}(x+\nu)-n_{\rho\sigma}(x)\Big],\nonumber\\
 n_{\rho\sigma}&=&-2,-1,0,1,2,
\end{eqnarray}
where the number of Dirac sheets $n_{\mu\nu}(x)$ is given by 
the difference
\begin{equation}
\frac{1}{2\pi}\Big[
\bar\Phi_{\mu\nu}(x)-
\Phi_{\mu\nu}(x)\Big]
\end{equation}
between physical flux $\bar\Phi_{\mu\nu}(x)$ and plaquette flux
$\Phi_{\mu\nu}(x)$\cite{DEGRAND}. For a link $U_{\rho}$, to be updated
in a boundary-cube with normalvector $e_{\mu}$, we search for the
plaquettes associated with the Dirac sheets $n_{\rho\sigma}(x)$ that
contribute to the monopole current $m_{\mu}(x)$ being parallel to
$e_{\mu}$. These plaquettes are depicted in Fig.~\ref{CUBE} with
dotted lines.
If, for any of the corresponding cubes, the change
in flux is larger than $2\pi$, the proposed link change is rejected in
the Monte Carlo.  This particular update rule fulfills {\em detailed
balance} and is equivalent to an infinitely large chemical potential
for the monopoles on the boundary shell\cite{BORNYAKOV} with the
action $~S = S_{W} + S_{\lambda}~$. $~S_{W}~$ is the standard Wilson
action and $~S_{\lambda}~$ is the additional term which suppresses
monopole currents $m_{\mu}$ at the lattice slice $x_{\mu}=0$:
\begin{equation}
S_{\lambda} = \lambda \cdot \sum_{\mu} \sum_{x:~x_{\mu}=0} \mid
m_{\mu} (x) \mid , ~~\lambda \rightarrow \infty .
\end{equation}
$S_{\lambda}$ explicitly breaks translation symmetry.
\section{RESULTS}

On the $16^4$- as well as on the $24^4$-lattice, we first scanned the
$\beta$-range in order to get a rough idea about the location of the
(pseudo)-critical region.  It is shifted towards smaller
$\beta$-values as expected from monopole suppression. Table~1 shows
the number of Metropolis sweeps we performed subsequently for the such
chosen $\beta$-values and the plaquette averaged over the whole
lattice.  On the $24^4$-lattice in the average about 1\% of the
proposed updates are rejected at the boundary.

The plaquettes have been sampled on the surfaces of sub-hypercubes
nested around the centre of the lattice.  The 3D projection of these
shells is shown in Fig.~\ref{NEST}.  The boundary shell is denoted by
the index $1$, the inner shell gets the index $L/2$.
\vglue12pt
\noindent
\begin{minipage}{7.5cm}
\newlength{\digitwidth} \settowidth{\digitwidth}{\rm 0}
\catcode`?=\active \def?{\kern\digitwidth}
\noindent
{Table 1: {Lattice sizes, $\beta$-values and \# of sweeps.}
\footnotesize
\begin{tabular*}{7.5cm}{@{}l@{\extracolsep{\fill}}rrrr}
\hline
size & $\beta$ & \# of sweeps & $<P>$ \\
\hline
$16^4$ & $0.990$   &30000  & 0.6290(2)\\
       & $0.995$   &10000  & 0.6408(4)\\
       & $1.000$   &60000  & 0.6530(2)\\
       & $1.002$   &60000  & 0.6584(1)\\
       & $1.003$   &60000  & 0.6599(1)\\
       & $1.005$   &60000  & 0.6630(1)\\
       & $1.010$   &10000  & 0.6688(2)\\
\hline
$24^4$ & $1.0060$  &60000  & 0.6546(1)\\
       & $1.0065$  &60000  & 0.6572(1)\\
       & $1.0070$  &250000 & 0.65856(5)\\
       & $1.007125$&200000 & 0.65935(9)\\
       & $1.0075$  &80000  & 0.66010(9)\\
       & $1.0078$  &50000  & 0.66099(6)\\
       & $1.0080$  &55000  & 0.66139(5)\\
\end{tabular*}}
\end{minipage}
\vglue 12pt
\begin{figure}
\epsfxsize=7.0 cm
\centerline{\epsfbox{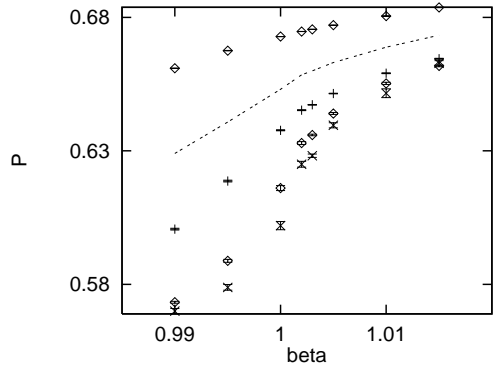}}
\vglue-12pt
\caption[a]{Average plaquette (dotted line) and plaquettes on 
shells 1,3,5 and 7 (from top to bottom) on the
$16^4$-lattice.\label{PLAQ16}}
\end{figure}
The boundary shell is always in a deconfined state. This is a well
known phenomenon in the absence of monopoles\cite{BORNYAKOV}.
Therefore, the variation of the plaquette value within the metastable
region is very small for this shell. As the boundary shell contains
the outweighting part of the sites, the plaquette average over the
whole lattice is dominated by the contribution of the boundary.  We
found the influence of the boundary decreasing towards the centre,
{\em i.e.}\ the plaquette values show increasing `steps' (as functions
of $\beta$) going to the inner shells, see Fig.~\ref{PLAQ16} for the
$16^4$-lattice.  For inner shells the plaquette values are close to
each other and their behaviour is more and more like that of the
plaquette of the $H_4$ system with p.b.c.  We interpret this results
as the onset of the recovery of translation invariance around the
centre.

We carefully analyzed the time series for the inner shells on the
$16^4$-lattice but the series merely exhibited a chaotic behaviour and
no metastabilities could be identified. Therefore we decided to 
work on a very large $24^4$-system and  we performed altogether nearly 
one million sweeps.
\begin{figure}
\epsfxsize=7.0 cm
\centerline{\epsfbox{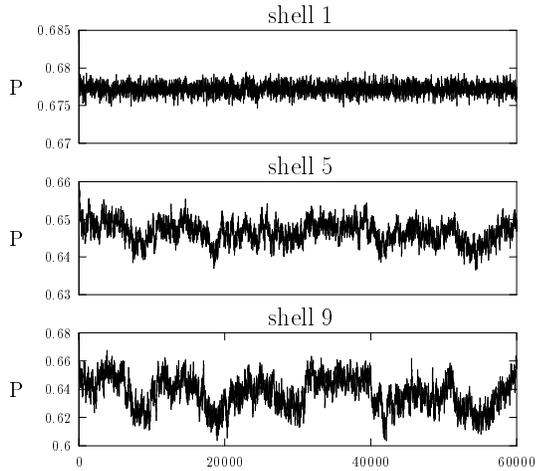}}
\vglue-12pt
\caption[a]{Time series on $24^4$-lattice for different shells.\label{SERIES24}}
\end{figure}

\begin{figure}
\epsfxsize=5.0 cm
\centerline{\epsfbox{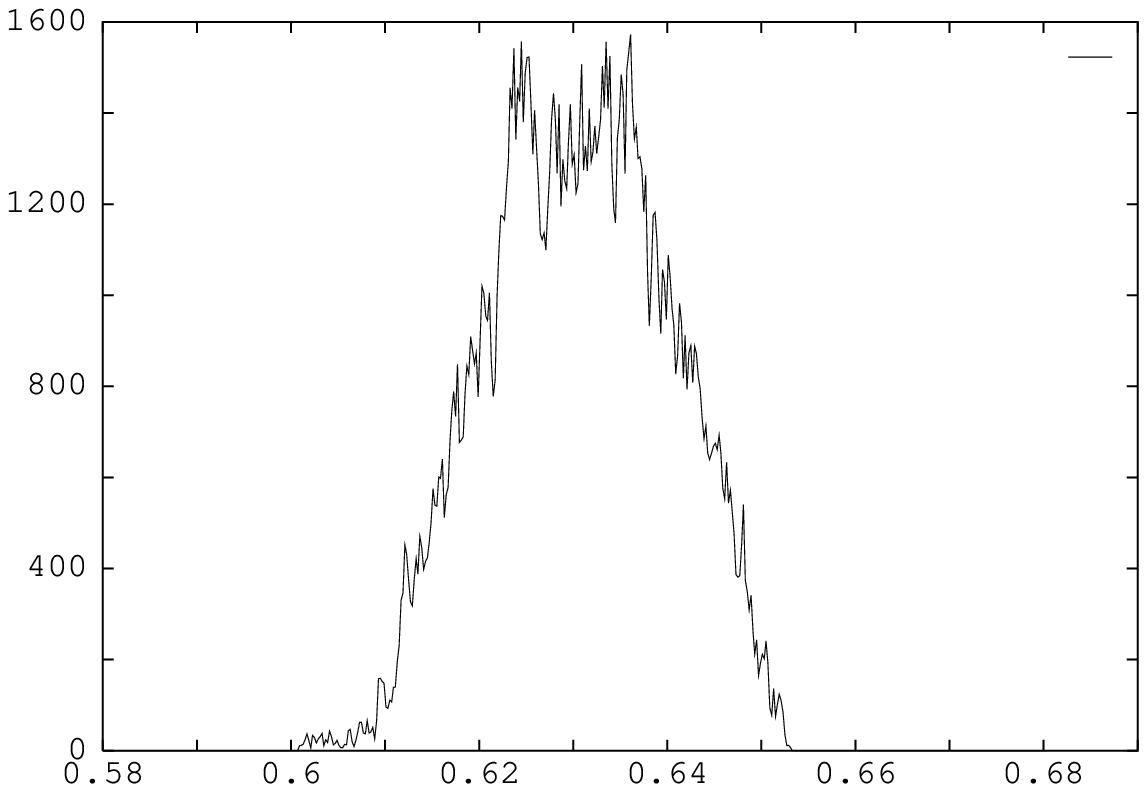}}
\centerline{$P$}
\vglue-18pt
\caption[a]{Histogram of shell 9 after removing of fluctuations.\label{DOUBLE}}
\end{figure}

On this lattice, we found a much broader region in 4-d space at the
centre where the behaviour of the system resembled that of U(1) on
$H_4$ with p.b.c.  and we found evidence for metastabilities in the
time series.  In Fig.~\ref{SERIES24}, the time series of plaquettes
for the shells 1, 5 and 9 are depicted for $\beta=1.007125$.  At this
value of $\beta$ we encountered long metastabilities extending over
more than 10000 sweeps as can be most clearly seen in the series of
shell 9.  As the fluctuations of the plaquette in inner shells are
larger than the gap between the two metastable states we performed a
windowing procedure in order to average out the fluctuations on small
scales and applied the method to the time series of shell 9. The
result is presented in Fig.~\ref{DOUBLE}, where we see a non-ambiguous
double-peak structure of the histogram.  More detailed results will be
presented in a forthcoming paper.

\section{CONCLUSIONS}
In our simulations  \\
\noindent
-- we did not observe first order signals on small lattices because of
boundary effects. We believe this is true for f.b.c. as well.\\
\noindent
-- we observed metastabilities coming up on the $24^4$-lattice. This
implies that periodically closed monopole loops can not be the reason
for the first order nature of U(1) with p.b.c.

\end{document}